\documentstyle[12pt]{article}
\def\ssl#1{\rlap{\hbox{$\mskip 3 mu /$}}#1}  
\begin{document}

\def\bigZ{Z\!\!\!Z}
\def\Nequalstwo{\Psi}
\def\eff{{\rm eff}}
\def\inst{{\rm inst}}
\def\fermi{{\rm fermi}}
\def\trtwo{\tr^{}_2\,}
\def\finv{f^{-1}}
\def\Ubar{\bar U}
\def\wbar{\bar w}
\def\fbar{\bar f}
\def\abar{\bar a}
\def\bbar{\bar b}
\def\Deltabar{\bar\Delta}
\def\dalpha{{\dot\alpha}}
\def\dbeta{{\dot\beta}}
\def\dgamma{{\dot\gamma}}
\def\ddelta{{\dot\delta}}
\def\Sbar{\bar S}
\def\Im{{\rm Im}}
\def\sst{\scriptscriptstyle}
\def\cld{C_{\sst\rm LD}^{}}
\def\csd{C_{\sst\rm SD}^{}}
\def\bigI{{\rm I}_{\sst 3\rm D}}
\def\Mr{{\rm M}_{\sst R}}
\def\cJ{C_{\sst J}}
\def\one{{\sst(1)}}
\def\two{{\sst(2)}}
\def\vsd{v^{\sst\rm SD}}
\def\vasd{v^{\sst\rm ASD}}
\def\Phibar{\bar\Phi}
\def\F{{\cal F}_{\sst\rm SW}}
\def\P{{\cal P}}
\def\A{{\cal A}}
\def\susy{supersymmetry}
\def\sigmabar{\bar\sigma}
\def\barsigma{\sigmabar}
\def\ASD{{\scriptscriptstyle\rm ASD}}
\def\cl{{\,\rm cl}}
\def\lambdabar{\bar\lambda}
\def\R{{R}}
\def\psibar{\bar\psi}
\def\sqrtwo{\sqrt{2}\,}
\def\etabar{\bar\eta}
\def\Thetabar{{\bar\Theta_0}}
\def\Qbar{\bar Q}
\def\susic{supersymmetric}
\def\vhiggs{{\rm v}}
\def\vhiggsa{{\cal A}_{\sst00}}
\def\vbarhiggs{\bar{\rm v}}
\def\vhiggsbar{\bar{\rm v}}
\def\novetal{Novikov et al.}
\def\Novetal{Novikov et al.}
\def\ADS{Affleck, Dine and Seiberg}
\def\ads{Affleck, Dine and Seiberg}
\def\setI{\{{\cal I}\}}
\def\Abar{A^\dagger}
\def\B{{\cal B}}
\def\infinity{\infty}
\def\C{{\cal C}}
\def\Psitwo{\Psi_{\scriptscriptstyle N=2}}
\def\Psibartwo{\bar\Psi_{\scriptscriptstyle N=2}}
\def\ms{Minkowski space}
\def\zero{{\scriptscriptstyle(0)}}
\def\new{{\scriptscriptstyle\rm new}}
\def\u{\underline}
\def\uA{\,\lower 1.2ex\hbox{$\sim$}\mkern-13.5mu A}
\def\uX{\,\lower 1.2ex\hbox{$\sim$}\mkern-13.5mu X}
\def\uD{\,\lower 1.2ex\hbox{$\sim$}\mkern-13.5mu {\rm D}}
\def\uDzero{{\uD}^\zero}
\def\uAzero{{\uA}^\zero}
\def\upsizero{{\upsi}^\zero}
\def\uF{\,\lower 1.2ex\hbox{$\sim$}\mkern-13.5mu F}
\def\uW{\,\lower 1.2ex\hbox{$\sim$}\mkern-13.5mu W}
\def\uWbar{\,\lower 1.2ex\hbox{$\sim$}\mkern-13.5mu {\overline W}}
\def\Dbar{D^\dagger}
\def\Fbar{F^\dagger}
\def\uAbar{{\uA}^\dagger}
\def\uAbarzero{{\uA}^{\dagger\zero}}
\def\uDbar{{\uD}^\dagger}
\def\uDbarzero{{\uD}^{\dagger\zero}}
\def\uFbar{{\uF}^\dagger}
\def\uFbarzero{{\uF}^{\dagger\zero}}
\def\uV{\,\lower 1.2ex\hbox{$\sim$}\mkern-13.5mu V}
\def\uZ{\,\lower 1.2ex\hbox{$\sim$}\mkern-13.5mu Z}
\def\uv{\lower 1.0ex\hbox{$\scriptstyle\sim$}\mkern-11.0mu v}
\def\uq{\lower 1.0ex\hbox{$\scriptstyle\sim$}\mkern-11.0mu q}
\def\uc{\lower 1.0ex\hbox{$\scriptstyle\sim$}\mkern-11.0mu c}
\def\uPsi{\,\lower 1.2ex\hbox{$\sim$}\mkern-13.5mu \Psi}
\def\uPhi{\,\lower 1.2ex\hbox{$\sim$}\mkern-13.5mu \Phi}
\def\uchi{\lower 1.5ex\hbox{$\sim$}\mkern-13.5mu \chi}
\def\uf{\lower 1.5ex\hbox{$\sim$}\mkern-13.5mu f}
\def\utheta{\lower 1.5ex\hbox{$\sim$}\mkern-13.5mu \theta}
\def\chitilde{\tilde \chi}
\def\etatilde{\tilde \eta}
\def\uchitilde{\lower 1.5ex\hbox{$\sim$}\mkern-13.5mu \tilde\chi}
\def\ueta{\lower 1.5ex\hbox{$\sim$}\mkern-13.5mu \eta}
\def\uetatilde{\lower 1.5ex\hbox{$\sim$}\mkern-13.5mu \tilde\eta}
\def\Psibar{\bar\Psi}
\def\uPsibar{\,\lower 1.2ex\hbox{$\sim$}\mkern-13.5mu \Psibar}
\def\upsi{\,\lower 1.5ex\hbox{$\sim$}\mkern-13.5mu \psi}
\def\uphi{\lower 1.5ex\hbox{$\sim$}\mkern-13.5mu \phi}
\def\psibar{\bar\psi}
\def\upsibar{\,\lower 1.5ex\hbox{$\sim$}\mkern-13.5mu \psibar}
\def\etabar{\bar\eta}
\def\uetabar{\,\lower 1.5ex\hbox{$\sim$}\mkern-13.5mu \etabar}
\def\chibar{\bar\chi}
\def\uchibar{\,\lower 1.5ex\hbox{$\sim$}\mkern-13.5mu \chibar}
\def\upsibarzero{\,\lower 1.5ex\hbox{$\sim$}\mkern-13.5mu \psibar^\zero}
\def\ulambda{\,\lower 1.2ex\hbox{$\sim$}\mkern-13.5mu \lambda}
\def\ulambdabar{\,\lower 1.2ex\hbox{$\sim$}\mkern-13.5mu \lambdabar}
\def\ulambdabarzero{\,\lower 1.2ex\hbox{$\sim$}\mkern-13.5mu \lambdabar^\zero}
\def\ulambdabarnew{\,\lower 1.2ex\hbox{$\sim$}\mkern-13.5mu \lambdabar^\new}
\def\D{{\cal D}}
\def\M{{\cal M}}
\def\N{{\cal N}}
\def\Dslash{\,\,{\raise.15ex\hbox{/}\mkern-12mu \D}}
\def\Dbarslash{\,\,{\raise.15ex\hbox{/}\mkern-12mu {\bar\D}}}
\def\delslash{\,\,{\raise.15ex\hbox{/}\mkern-9mu \partial}}
\def\delbarslash{\,\,{\raise.15ex\hbox{/}\mkern-9mu {\bar\partial}}}
\def\L{{\cal L}}
\def\hf{{\textstyle{1\over2}}}
\def\quarter{{\textstyle{1\over4}}}
\def\twe{{\textstyle{1\over12}}}
\def\eighth{{\textstyle{1\over8}}}
\def\fourth{\quarter}
\def\wb{Wess and Bagger}
\def\xibar{\bar\xi}
\def\ss{{\scriptscriptstyle\rm ss}}
\def\sc{{\scriptscriptstyle\rm sc}}
\def\uvcl{{\uv}^\cl}
\def\uAcl{\,\lower 1.2ex\hbox{$\sim$}\mkern-13.5mu A^{}_{\cl}}
\def\uAbarcl{\,\lower 1.2ex\hbox{$\sim$}\mkern-13.5mu A_{\cl}^\dagger}
\def\upsinew{{\upsi}^\new}
\def\ASDzero{{{\scriptscriptstyle\rm ASD}\zero}}
\def\SDzero{{{\scriptscriptstyle\rm SD}\zero}}
\def\SD{{\scriptscriptstyle\rm SD}}
\def\varthetabar{{\bar\vartheta}}
\def\three{{\scriptscriptstyle(3)}}
\def\dagthree{{\dagger\scriptscriptstyle(3)}}
\def\ld{{\scriptscriptstyle\rm LD}}
\def\vld{v^\ld}
\def\Dld{{\rm D}^\ld}
\def\Fld{F^\ld}
\def\Ald{A^\ld}
\def\Fbarld{F^{\dagger\scriptscriptstyle\rm LD}}
\def\Abarld{A^{\dagger\scriptscriptstyle \rm LD}}
\def\lambdald{\lambda^\ld}
\def\lambdabarld{\bar\lambda^\ld}
\def\psild{\psi^\ld}
\def\psibarld{\bar\psi^\ld}
\def\dsiginst{d\sigma_{\scriptscriptstyle\rm inst}}
\def\xione{\xi_1}
\def\xionebar{\bar\xi_1}
\def\xitwo{\xi_2}
\def\xitwobar{\bar\xi_2}
\def\thetatwo{\vartheta_2}
\def\thetatwobar{\bar\vartheta_2}
\def\Ltwo{\L_{\sst SU(2)}}
\def\Leff{\L_{\rm eff}}
\def\Laux{\L_{\rm aux}}
\def\oneloop{{\sst\rm 1\hbox{-}\sst\rm loop}}
\def\LSUtwo{{\cal L}_{\rm SU(2)}}
\def\Dhat{\hat\D}
\def\bkgd{{\sst\rm bkgd}}
\def\Lgft{{\cal L}_{\sst\rm g.f.t.}}
\def\Lghost{{\cal L}_{\sst\rm ghost}}
\def\Sinst{S_{\rm inst}}
\def\etal{{\rm et al.}}
\def\S{{\cal S}}

\newcommand{\nd}[1]{/\hspace{-0.5em} #1}
\begin{titlepage}
\begin{flushright}
NI-97020\\
SWAT-154 \\
April 1997 \\
hep-th/9704197
\end{flushright}
\begin{centering}
\vspace{.2in}
{\large {\bf Multi-Instantons, Three-Dimensional Gauge Theory, 
\\ and the Gauss-Bonnet-Chern Theorem }}\\
\vspace{.4in}
 N. Dorey$^{1}$, V. V. Khoze$^{2}$, M. P. Mattis$^{3}$ \\
\vspace{.4in}
$^{1}$ Physics Department, University of Wales Swansea \\
Singleton Park, Swansea, SA2 8PP, UK\\
\vspace{.3in}
$^{2}$ Physics Department, Centre for Particle Theory  \\
University of Durham, Durham DH1 3LE, UK \\
\vspace{.3in} 
$^{3}$ Theoretical Division,  Los Alamos National Laboratory \\
Los Alamos, NM 87545, USA\\
\vspace{.4in}
{\bf Abstract} \\
\end{centering}
We calculate multi-instanton effects in a three-dimensional gauge
theory with $N=8$ supersymmetry and gauge group $SU(2)$. The
$k$-instanton contribution to an eight-fermion correlator 
is found to be proportional to the Gauss-Bonnet-Chern integral
of the Gaussian curvature over the centered moduli space of charge-$k$ BPS
monopoles, $\tilde{\cal M}_{k}$. For $k=2$ the integral can be
evaluated using the explicit metric on $\tilde{\cal M}_{2}$ found by
Atiyah and Hitchin. In this case the integral is equal to the Euler
character of the manifold. 
More generally the integral is the volume contribution to the
index of the Euler operator on $\tilde{\cal M}_{k}$, which may differ from the
Euler character by a boundary term. 
We conjecture that the boundary terms vanish and evaluate  
the multi-instanton contributions using recent results for the
cohomology of $\tilde{\cal M}_{k}$. We comment briefly on the
implications of our result for a recently proposed test of 
M(atrix) theory.


\end{titlepage}

The important recent advances in understanding the low-energy dynamics
of supersymmetric gauge theory have led to some interesting exact
results in three spacetime dimensions (3D). In
particular, exact metrics have been proposed for both Coulomb and Higgs
branches of the $N=4$ theories \cite{seib}-\cite{ber2} 
while exact superpotentials have been 
determined for the $N=2$ theories \cite{ne2ber}\cite{ne2s}. 
Surprisingly, these results can be
seen as consequences of non-perturbative dualities in string theory,
where the 3D SUSY field theories appear on 
the world volume of membranes. Very recently, an interesting
application has also been proposed \cite{polch}
for $N=8$ SUSY gauge theory in
three dimensions as a description of membrane scattering in 
M(atrix) theory. 
\paragraph{}
In all these cases, an important role is played by instanton
effects in the three-dimensional gauge theory
\cite{SW3}\cite{dkmtv}\cite{polch}. In fact, the exact 
results yield predictions for all instanton contributions to the
low-energy theory. As in four-dimensions 
\cite{finnell}-\cite{ao},  
first-principles semiclassical
calculations of these effects yield independent quantitative 
tests of the proposed exact results 
\cite{sw12} and therefore of the duality on which they 
rely. In recent work \cite{dkmtv}, 
we calculated a one-instanton contribution to the low-energy effective
action in the three-dimensional
$N=4$ theory with gauge group $SU(2)$. This result verifies the
conjecture of Seiberg and Witten \cite{SW3}
that the Coulomb branch of this
theory is isometric to the Atiyah-Hitchin manifold. 
\paragraph{}
In this paper we will focus on the $N=8$ theory.
Polchinski and Pouliot \cite{polch} have 
calculated a one-instanton contribution 
in this theory and compared it with a certain scattering
amplitude for membranes in eleven-dimensional supergravity. 
In general, the $k$-instanton contributions in the  $N=8$ three-dimensional
theory can be thought of as an approximation 
to the M(atrix) theory scattering amplitude of two supermembranes
with $k$ units of the M-momentum transfer. The $k$-instanton result
can then be compared to an independent calculation of the membrane
scattering in eleven-dimensional supergravity. The authors of
\cite{polch} found agreement between the result of an explicit 
one-instanton calculation and the corresponding supergravity scattering
amplitude for a single unit of M-momentum transfer.
This supports the conjectured exactness of the matrix model
description of M-theory. In this paper we give the
multi-instanton generalization of this result. As shown below,
for all values of $k$, multi-instantons correctly
reproduce the M-momentum dependence 
of the supergravity amplitude calculated in \cite{polch}, 
up to perturbative corrections in the multi-instanton background which
we have not calculated. 
\paragraph{}
In the three-dimensional gauge theories considered here, 
the relevant instanton configurations are BPS monopoles. 
At weak-coupling, path-integration in the sector of topological charge
$k$ reduces to a finite-dimensional integral over the moduli-space of
$k$ BPS monopoles. In a supersymmetric theory, it is also necessary to
integrate over Grassmann parameters corresponding to 
the fermionic zero-modes of the instanton. For the $N=8$ theory, we will
show that all but eight of these modes are lifted by a four-fermion 
term in the classical action of the supersymmetric instanton. This
means that there are non-zero corrections 
to the eight-fermion vertex calculated in \cite{polch} from all
numbers of instantons. Our main result is that the $k$-instanton 
contribution is given by the Gauss-Bonnet-Chern (GBC) integral of the
Gaussian curvature on the centered moduli-space $\tilde{\cal M}_{k}$.         
This integral is the volume contribtion to the index of the Euler
operator on $\tilde{\cal M}_{k}$. 
For $k=2$, the GBC integral has been evaluated 
by Gauntlett and Harvey \cite{gh}, 
using the explicit metric on the two-monopole moduli
space obtained by Atiyah and Hitchin \cite{AH}. In this case the volume
contribution is exactly equal to the Euler character of $\tilde{\cal
M}_{2}$ which agrees with the earlier conclusion 
\cite{GPR} that the boundary 
terms for the relevant index theorem for any 
Asymptotically Locally Flat (ALF) metric should vanish. 
We propose the extension of this equality for all $k$. The
cohomology of the higher charge monopole moduli-spaces has recently been
determined by Segal and Selby
\cite{segal}.  The Euler characters of these spaces
then yield the multi-instanton contributions. It is notable that these
contributions, which at first sight have a highly
non-trivial dependence on the unknown hyper-K\"{a}hler metric on 
$\tilde{\cal M}_{k}$, in fact turn out to be topological invariants of
these manifolds. 
\paragraph{}
Theories with extended SUSY in three spacetime dimensions are  
straightforwardly obtained from theories with the same number of
component supercharges in four-dimensions (4D) by
dimensional reduction. In the following the integer $N$ denotes the
number of real two-component Majorana supercharges of the theory in question. 
Following \cite{hanwit}, 
we will also define ${\cal N}=N/2$ which counts the corresponding number of 
Weyl supercharges for a 4D theory.   
In Ref.~\cite{dkmtv} 
we discussed the $N=4$ theory in
3D which is the dimensional reduction of the minimal ${\cal N}=2$ theory in
4D. The latter theory consists of an ${\cal N}=2$ gauge multiplet which
contains the (4D) gauge field $\uv_{m}$ ($m=0,1,2,3$), a complex scalar 
$\uA$ and two species of Weyl
fermion $\ulambda_{\alpha}$ and $\upsi_{\alpha}$ all in the adjoint
representation of the gauge group.\footnote{
Our conventions are the same as in \cite{dkmtv},\cite{dkm1}: in
particular we use undertwiddling 
for the fields in the $SU(2)$ matrix notation,
$\uX \equiv X^a\tau^a/2$.}
In the following we will be interested in a
three-dimensional theory with twice as many supercharges which can be
constructed by dimensionally reducing the four-dimensional ${\cal N}=4$
theory \cite{sb}. This four-dimensional theory is obtained by augmenting the
gauge multiplet introduced above with an adjoint ${\cal N}=2$ hypermultiplet
which contains two complex scalars $\uq$ and $\tilde{\uq}$ and their
Weyl fermion superpartners $\uf_{\alpha}$ and
$\tilde{\uf_{\alpha}}$. The resulting supersymmetry algebra has an
$SU(4)_{\cal R}$ group of automorphisms and 
it is convenient to rewrite the four species of Weyl fermions 
$\{\ulambda_\alpha,\upsi_\alpha,\uf_\alpha,\tilde{\uf_\alpha}\}$
 as $\ulambda_{\alpha}^{M}$ where  $M=1,2,3,4$.
\paragraph{}
In the following we will compactify one of the four
spatial dimensions, say
$x_{3}$, 
on a circle
of radius $R$. We will be interested in
the three-dimensional quantum theory obtained by integrating only over
field configurations which are independent of the compactified
coordinate. In this case we may integrate over $x_{3}$ in the action
to obtain, 
\begin{eqnarray}
\frac{1}{g^{2}}\int\, d^{4}x & \rightarrow &  \frac{2\pi}{e^{2}}\int\, d^{3}x
\label{coupling}
\end{eqnarray}
where $e=g/\sqrt{R}$ defines the dimensionful 3D gauge coupling in
terms of the dimensionless 4D counterpart $g$. 
The resulting $N=8$ supersymmetric gauge theory in three-dimensions
contains seven real scalar fields and a (3D) gauge field as well as
eight real Majorana fermions. The Lagrangian has a $spin(7)$ 
${\cal R}$-symmetry group which contains both the $SU(4)_{\cal R}$
symmetry of the 4D theory as well as a new symmetry group $SU(2)_{N}$
which appears in 3D \cite{sb}. In the following we will not work with the
three-dimensional fields directly.    
In considering instanton effects it will instead 
be convenient to use the (dimensionally-reduced) fields of the 4D
theory introduced above. 
\paragraph{}
The theory has a Coulomb branch on which the adjoint scalar fields
acquire mutually commuting expectation values. In this phase, two
components of the gauge field gain a mass, $M_{W}$, by the adjoint
Higgs mechanism. The remaining massless photon is eliminated in favour
of a periodic scalar $0\leq \sigma <2\pi$ via a duality
transformation. The resulting classical 
moduli space is a flat eight-dimensional
manifold parametrized by the massless components of the seven adjoint
scalars and the dual photon. $N=8$ supersymmetry forbids quantum
corrections to the metric on the Coulomb branch, hence the 
low-energy effective action with up to two derivatives or 
four fermions is a free field theory.
 In fact the first non-trivial quantum corrections
appear at the fourth order in the derivative expansion and
correspondingly we will calculate corrections to a correlator with eight
fermion insertions.  
\paragraph{}
Three-dimensional gauge theories with extended supersymmetry have
instanton solutions: the relevant field
configurations are BPS multi-monopoles of magnetic charge $k$. 
These configurations have finite Euclidean action equal to
$|k|S_{0}-ik\sigma$ where $S_{0}=(8\pi^{2}M_{W})/e^{2}$ and the
imaginary term involving the dual photon comes from a surface term
analagous to the $\theta$-term in four dimensions. To describe the
zero modes of these supersymmetric instantons it is convenient to
make a particular vacuum choice \cite{dkmtv}: the component, 
$\uv_{3}$, of the four-dimensional gauge
field (which is a scalar in 3D) 
acquires a non-zero expectation value, as does $\sigma$;  all other
VEVs are set to zero. As usual we choose the non-zero VEV to lie in
the third direction in the $SU(2)$ gauge group: $\langle \uv_{3}
\rangle= \sqrt{2}{\rm v} \tau^{3}/2$ where ${\rm v}$ is real and positive and 
$M_{W}=\sqrt{2}{\rm v}$. In this case the static Bogomol'nyi 
equation satisfied by the gauge and Higgs components of the BPS monopole 
of charge $k>0$ can be rewritten as a self-dual
Yang-Mills (SDYM) equation for the four-dimensional gauge field, $\uv_{m}$ 
\cite{lohe}:
\begin{equation}
\uv^{\rm cl}_{mn}\ =\ ^{*}\uv^{\rm cl}_{mn}  
\label{SD}
\end{equation}    
Solutions of negative magnetic charge are obtained from the corresponding
anti-self-dual equation. 
\paragraph{}
In order to find the zero modes, $\delta\uv^{m}=\uZ^{m}$ of the
multi-monopole solution $\uv^{\rm cl}_{m}$ it is necessary to solve
the linearized SDYM equation together with a background gauge
condition
\begin{equation}
{\cal D}_{\rm cl}^{[m}{\uZ}^{n]}=\,^{*}{\cal D}_{\rm cl}^{[m}{\uZ}^{n]}
\ \ \ ,\ \ \ {\cal D}_{{\rm cl}}^{m}{\uZ}^{m}=0 \ .
\label{fan}
\end{equation}
where ${\cal D}^{m}_{\rm cl}$ is the adjoint gauge-covariant derivative 
in the self-dual gauge background. The Callias index theorem
\cite{cal}\cite{w1} 
tells us
that there are exactly $4k$ normalizable solutions of these equations
which we denote $\uZ^{(i)}_{m}$ for $i=1,2\ldots 4k$. Correspondingly 
we can introduce collective coordinates $X_{i}$ for the
multi-monopole solutions which (locally) parametrize a 
charge-$k$ moduli-space, ${\cal
M}_{k}$. The properties of these spaces have been extensively
studied by mathematicians and are described in detail in \cite{AH}. 
They are smooth Riemannian manifolds
equipped with a natural metric, 
\begin{equation}
g_{ij}=\frac{2\pi}{e^{2}}\int d^{3}x\,2{\rm Tr}\,\uZ_{m}^{(i)} 
\uZ_{m}^{(j)}
\label{metric}
\end{equation}
In addition the monopole moduli-spaces inherit three inequivalent
complex structures from the three inequivalent self-dual complex structures on
$R^{4}$. These complex structures are covariantly constant with respect
to the metric (\ref{metric}) and generate a representation of the
quaternions on the tangent bundle of ${\cal M}_{k}$. The
$k$-monopole moduli space is therefore a hyper-K\"{a}hler manifold of
real dimension $4k$.    
\paragraph{}
Spatial translations act freely on
${\cal M}_{k}$, while global rotations in the unbroken 
$U(1)$ subgroup of the gauge group act with a ${\bigZ}_{k}$
stabilizer. The moduli space can be isometrically decomposed as, 
\begin{eqnarray}
{\cal M}_{k} & \simeq  & 
R^{3}\times \frac{S^{1}\times \tilde{\cal M}_{k}}{{\bigZ}_{k}}    
\label{decompose}
\end{eqnarray}
The $R^{3}$ and $S^{1}$ factors are parametrized by the 3D spacetime position
vector $X_{\mu}$, $\mu=0,1,2$ and overall charge angle $X_{3}=\theta$ 
of the $k$-monopole solution 
respectively. The corresponding metric $\bar{g}_{mn}$, 
obtained by restriction of $g_{ij}$ to $R^{3}\times S^{1}$, 
is flat and can be written as,  
\begin{eqnarray}
\bar{g}_{mn} & = & \delta_{mn} g^{}_{XX}+\delta_{m3}\delta_{n3}
\left(g^{}_{\theta\theta}-g^{}_{XX}\right) \ , \qquad m,n=0,1,2,3
\label{rtimess1}
\end{eqnarray}
\paragraph{}
In order to calculate the instanton contibution to the Euclidean correlators of
the theory, it is necessary to integrate over the collective
coordinates with the measure obtained from changing variables in the
path integral. The contribution of the four global symmetry modes to
the measure can be written as \cite{Bernard},   
\begin{eqnarray}
\int\,d\bar{\mu}_{B}  & =& \int\,\frac{d^{3}X}{(2\pi)^{\frac{3}{2}}} 
(g^{}_{XX})^{\frac{3}{2}} \int_{0}^{\frac{2\pi}{k}}\,\frac{d{\theta}}
{(2\pi)^{\frac{1}{2}}} (g^{}_{\theta\theta})^{\frac{1}{2}}
\label{bmeasure}
\end{eqnarray} 
where the limit of integration on the $\theta$-integral reflects
the discrete ${\bigZ}_{k}$ symmetry. The overall normalization of the 
translational and charge rotation zero modes of a single monopole
were given in Appendix C of our previous paper \cite{dkmtv}. The 
results generalize trivially to all $k$ giving $g^{}_{XX}=kS_{0}$ and
$g^{}_{\theta\theta}=kS_{0}/M_{W}^{2}$. These constants are related to
the mass and moment of inertia of the $k$-monopole solution
respectively. 
\paragraph{}
The remaining factor in (\ref{decompose}) is the reduced or 
centered moduli space $\tilde{\cal M}_{k}$. This $4(k-1)$ dimensional 
hyper-K\"{a}hler manifold is the
simply-connected $k$-fold cover of the moduli-space of $k$ monopoles
with a fixed centre of mass and global $U(1)$ phase. $\tilde{\cal M}_{k}$ is 
parametrized by coordinates $Y_{q}$, with 
$q=1,2,\ldots 4(k-1)$. In an asymptotic regime, 
these parameters can be identified as the 
relative separations and charge angles of $k$ distinct BPS monopoles.
We will write the restriction of the metric $g_{ij}$ on ${\cal M}_{k}$
to the reduced moduli space, $\tilde{\cal M}_{k}$, as
$\tilde{g}_{pq}$. The corresponding contribution to the path integral
measure can be expressed as an integral over $\tilde{\cal M}_{k}$, 
\begin{eqnarray}
\int\,d\tilde{\mu}_{B} & = & \int \ 
\frac{\prod_{q=1}^{4(k-1)}dY_{q}}{(2\pi)^{2(k-1)}}\, \sqrt{{\rm det}
\left(\tilde{g}\right)}
\label{relmeasure}
\end{eqnarray}   
The metric $\tilde{g}_{pq}$ is known exactly only in the two-monopole case 
where it was constructed explicitly by Atiyah and Hitchin
\cite{AH}. For $k>2$, only the asymptotic form of the metric in the
limit of well-separated monopoles is known \cite{gm2}. Fortunately our
final result will not depend on the metric explicitly but only on the
global structure of $\tilde{\cal M}_{k}$. 
\paragraph{}
The problem of finding zero modes of the fermion fields in the
monopole background is closely related to the corresponding bosonic
problem described above. As in \cite{dkmtv}, it is convenient to
consider the classical equations of motion for the
dimensionally-reduced Weyl fermions of the 4D theory. 
\begin{eqnarray}
{\ssl \bar{\cal D}_{\rm cl}^{\dot{\alpha}\alpha}}\ulambda^{\rm cl}_{\alpha} & =
& 0 \label{dbar} \\
{\ssl {\cal D}}_{\rm cl}^{\alpha\dot{\alpha}}
\bar{\ulambda}^{\rm cl}_{\dot{\alpha}} & =
& 0 
\label{d}
\end{eqnarray}
In a monopole background of positive magnetic charge, 
Eq.~(\ref{dbar}) has normalizable solutions while 
Eq.~(\ref{d}) has none. In fact there is a simple relation between the
zero modes of ${\ssl \bar{\cal D}}_{\rm cl}$ and the zero modes of the
corresponding linearized equation (\ref{fan}) for 
the gauge field. For each bosonic zero mode, $\uZ_{m}^{(i)}$, the
spinor $\ulambda_{\alpha}^{(i,\dot{\alpha})}=
\sigma_{\alpha\dot{\alpha}}^{m}\uZ^{m\,(i)}$
with $\dot{\alpha}={1}$ and ${2}$, yields two
linearly-independent 
solutions for the adjoint Dirac equation (\ref{dbar}). 
\paragraph{}
Naively, the above argument indicates that each species of Weyl
fermion has $8k$ zero modes. However, as
explained in \cite{gaunt}\cite{blum}, this overcounts the number of
linearly independent solutions by a factor of four. The essential observation
is that the three inequivalent complex structures on the moduli space, 
$(J^a)_i^{\ j}$ for $a=1,2,3$, act on the vector zero modes as, 
\begin{eqnarray}  
(J^a)_i^{\ j}\uZ_{m}^{(i)} & = & {\eta}^{a}_{mn}\uZ_{n}^{(j)}
\label{cstructure}
\end{eqnarray}
where ${\eta}^{a}_{mn}$ are the self-dual generators of
$SO(4)$ rotations of the 4D vector index $m$. It then follows easily
that (for fixed indices $j$ and $\alpha$) the six zero modes
$(J^a)_i^{\ j}
\ulambda_{\alpha}^{(i,\dot{\alpha})}$, for 
$a=1,2,3$  and $\dot{\alpha}={1},{2}$, 
are linear combinations of the two independent 
modes $\ulambda_{\alpha}^{(j,\dot{\alpha})}$.        
Eliminating this overcounting, the 
total number of normalizable zero modes of the Dirac
operator in the charge-$k$ monopole background is $2k$ 
for each species of Weyl fermion. 
\paragraph{}
Just as for the bosonic zero modes, it is
convenient to consider separately 
the fermionic modes which correspond to the action of symmetry
generators on the monopole configuration. The bosonic fields of the
BPS monopole are invariant under half the SUSY generators. The action of
the remaining generators yields a total of eight zero modes of the
left-handed Weyl fermion fields. These can parametrized by four
Grassmann spinor collective coordinates, $\xi^{M}_{\beta}$ with
$M=1,2,3,4$, as             
\begin{eqnarray}
\ulambda^{{\rm cl}\, M}_{\alpha} & =& \hf\xi^{M}_{\beta}
(\sigma^{m}\sigmabar^{n})_{\alpha}^{\ \beta}
\uv^{\rm cl}_{mn} 
\label{fmodes}
\end{eqnarray}
The corresponding contribution to the multi-instanton measure is, 
\begin{equation}
\int \, d\bar{\mu}_{F}=\int\, \prod_{M=1}^{4} d^{2}\xi_{M}\, 
(k{\cal J}_{\xi})^{-4}
\label{fmeasure}
\end{equation}
where the normalization constant ${\cal J}_{\xi}$ is independent of
$k$. In Appendix C of \cite{dkmtv}, we obtained 
${\cal J}_{\xi}=2S_{0}$. The modes (\ref{fmodes}) 
are protected by $N=8$ SUSY and cannot be lifted. Hence to obtain a
non-zero contribution to the path integral,  
it will be necessary to saturate the Grassmann integrations appearing
in (\ref{fmeasure}) by inserting at least eight fermion fields (or
alternatively four fermion-bilinear parts of bose fields).  
\paragraph{}
The $8(k-1)$ remaining fermion zero 
modes do not correspond to the action of any
symmetries and their explicit forms are not known. In the following it
will suffice to relate these modes to their bosonic counterparts,
$\uZ_{m}^{(q)}$, where as before $q=1,2,\ldots 4(k-1)$. We will use
the fact that the tangent space $T_{k}$ at any point on $\tilde{\cal
M}_{k}$  is a vector field over the quaternions \cite{AH}. The three
inequivalent complex structures can then be represented as the unit
quaternions which act on the tangent vectors by quaternionic multiplication. 
Equivalently we can find a real basis for $T_{k}$ in which the zero modes
$\uZ_{m}^{(q)}$ (where $q=1,2,\ldots 4(k-1)$) 
can be partitioned into $k-1$ blocks of four 
on which the $su(2)$ algebra generated by
the complex structures acts irreducibly. Specifically 
we will choose a basis so that, for each value $M=1,2,3,4$, 
the index $q$ picks out one zero mode 
from each quaternionic block of four as it runs from $1$ to $4(k-1)$ over all
values satisfying $q=M \ {\rm mod}\,4$. A simple consequence is that, 
as $q$ runs over such a set of $k-1$ values, 
$\ulambda_{\alpha}^{(q,\dot{\alpha})}$ (with
$\dot{\alpha}={1}$ and ${2}$) form a set of $2(k-1)$
linearly independent fermion zero modes. 
It is convenient to parametrize the resulting set of $8(k-1)$ zero modes
of the four species of 4D Weyl fermions, 
$\ulambda^{M}_{\alpha}$, in terms of $4(k-1)$ Grassmann 
2-component real spinor
parameters $\alpha^{q}_{\beta}$ as       
\begin{eqnarray}
\ulambda^{{\rm cl}\, M}_{\alpha} & =& \sum_{q=M\ {\rm mod} \, 4} \,
\epsilon^{\beta\dot{\alpha}}
\alpha^{q}_{\beta}\ulambda_{\alpha}^{(q,\dot{\alpha})}
\label{fmodes2}
\end{eqnarray}
The corresponding contribution to the multi-instanton measure then has
the simple form, 
\begin{equation}
\int \, d\tilde{\mu}_{F}=\int\, \frac{ \prod_{q=1}^{4(k-1)}
d\alpha^{q}_{1} d\alpha^{q}_{2}   }
{{\rm det}\left(\tilde{g}\right)} 
\label{fmeasure2}
\end{equation}
\paragraph{}
As the modes appearing in (\ref{fmeasure2}) are not protected by any 
(super)symmetry it is natural to expect that they will be lifted. In
fact, in the corresponding instanton calculation in the
four-dimensional theory one finds 
\cite{dkm6}, 
that this lifting occurs precisely
due to Grassmann quadrilinear terms in the classical action of the
instanton. It is straightforward to show that similar terms
arise in the present case. By an explicit calculation  we find, 
\begin{equation}
S^{(k)}_{\rm quad}=\frac{1}{4}\tilde{R}_{pqrs} \alpha^{p}_1 \alpha^{q}_1
\alpha^{r}_2 \alpha^{s}_2
\label{quads}
\end{equation}
where $\tilde{R}_{pqrs}$ is the Riemann tensor formed from the
metric $\tilde{g}_{pq}$ on $\tilde{\cal M}_{k}$.  In particular one
can show that the vertex (\ref{quads}) is invariant under the
transformations of the collective coordinates which are induced by
$N=8$ supersymmetry transformations on the fields. Similarly
(\ref{quads}) is invariant under the abelian ${\cal R}$-symmetry, 
denoted $U(1)_{N}$ in \cite{SW3}, which prevents a similar lifting occuring in
the $N=4$ theory.   
\paragraph{}
A simple way to derive $S^{(k)}_{\rm quad}$ is to view the BPS
monopoles which are instantons in $R^{3}$ as solitons in 
$R^{3}\times S^{1}$, by introducing a compactified Euclidean `time' dimension,
$\tau$, $\ 0<\tau<\beta$. The $N=8$ SUSY Yang-Mills theory on $R^{3}$ is the 
$\beta \to 0$ limit of the corresponding theory on $R^{3}\times S^{1}$
with periodic boundary conditions in $\tau$ for fermions as well as bosons.
To correctly include leading order semiclassical effects, 
the monopole solutions must now be made 
`time'-dependent by allowing their bosonic and fermionic collective
coordinates to depend of $\tau$. After substituting such 
$\tau$-dependent $k$-monopole configurations into 
the action and integrating over $R^3$ one obtains an effective action
describing $\tau$-dependent geodesic motion of $k$ monopoles on the reduced
moduli space 
$\tilde{\cal M}_{k}$ with the metric $\tilde{g}_{pq}$, \cite{gaunt}\cite{blum}
\begin{equation}
S^{(k)}_{\rm eff}= kS_0 + \ \int_0^{\beta} d\tau \ \frac{1}{2}\left[
\ \tilde{g}_{pq} \ d_{\tau} Y^p d_{\tau} Y^q +
\ \tilde{g}_{pq} \ i \bar{\alpha}^p \gamma^0 D_{\tau} \alpha^q +
\ \frac{1}{12}\tilde{R}_{pqrs} (\bar{\alpha}^{p}  \alpha^{r})
(\bar{\alpha}^{q}  \alpha^{s})\right]
\label{meff}
\end{equation}
where $\bar{\alpha} = \alpha \gamma^0$ with $\gamma^0=\sigma^2$, and
$D_{\tau}\alpha^q = d_{\tau}\alpha^q + 
d_{\tau} Y^r \tilde{\Gamma}^p_{rq}\alpha^q $
is the covariant derivative on $\tilde{\cal M}_{k}$ formed from the
metric $\tilde{g}_{pq}$. 
Equation (\ref{meff}) is the action of an $N=8\times (1/2)$
supersymmetric Euclidean quantum mechanics\footnote{As explained
above, $N$ counts the number of two-component 
Majorana spinor supercharges. 
The notation $N=8\times(1/2)$ indicates that the SUSY QM described by 
(\ref{meff}) has eight one-component supercharges} on $\tilde{\cal M}_{k}$. 
The Riemann tensor term arises in (\ref{meff}) naturally
as the supersymmetric 
completion\footnote{In the case of ${\cal N}=2$ SUSY in 4D considered 
in \cite{gaunt}, 
the corresponding $S^{(k)}_{\rm eff}$ is an $N=4\times(1/2)$ quantum mechanics
on the moduli space and 
the Riemann tensor term is not generated.} of the kinetic terms 
in $S^{(k)}_{\rm eff}$. 
We can now dimensionally 
reduce and consider the $\tau$-independent contributions to 
$S^{(k)}_{\rm eff}$. In this case only the four-fermion term survives
and, rearranging the spinor indices, we obtain Eq. (\ref{quads}). 
An equivalent procedure is to let $\beta \to 0$
as in Ref.~\cite{AG}, after rescaling constant fermionic
coordinates by a factor of $\beta^{-1/4}$. 
In fact, as we will see below, the instanton contribution is formally 
equal to the regularized Witten index, ${\rm Tr}[(-1)^{F}\exp(-\beta
H)]$, of the quantum mechanics defined by the action (\ref{meff}) and
is therefore independent of $\beta$. 
\paragraph{}
As in any semiclassical instanton calculation, to complete the
specification of the measure we must also consider the contribution of
the non-zero modes. Usually in supersymmetric gauge theories these
contributions cancel between bose and fermi degrees of
freedom. However, in our recent work on $N=4$ SUSY gauge theory in
three dimensions \cite{dkmtv}, we found that this cancellation is not complete
due to the spectral asymmetry of the Dirac operator in the monopole
background. Specifically we found that the contribution of the gauge multiplet
(including ghosts) was equal to a non-trivial ratio of functional
determinants,
\begin{equation}
R=\left[\frac{{\rm det}(\Delta_{+})}{{\rm det'}(\Delta_{-})}
\right]^{\frac{1}{2}}
\label{cancel}
\end{equation} 
where $\Delta_{+}=\ssl{\bar{\cal D}_{\rm cl}}\ssl{\cal D}_{\rm cl}$,  
$\Delta_{-}=\ssl{{\cal D}}_{\rm cl}\ssl{\bar{\cal D}}_{\rm cl}$ and
${\rm det'}$ denotes the removal of zero eigenvalues. 
In the present case of $N=8$ SUSY, we must also include the
contribution of the adjoint hypermultiplet. A straightforward
calculation shows that this contribution is equal to $R^{-1}$. Hence,
the extra supersymmetries of the present case lead to a complete
cancellation of the non-zero mode contributions. Collecting all the
factors together, the complete multi-instanton measure can be written
as, 
\begin{eqnarray}
\int \, d\mu^{(k)} & = & 
\int \, d\bar{\mu}_{B}\,d\tilde{\mu}_{B} \ \int\,
d\bar{\mu}_{F}\,d\tilde{\mu}_{F} \ 
\exp\left(-kS_{0}+ik\sigma-S^{(k)}_{\rm quad}\right) \nonumber \\ 
& =& \int \, d\bar{\mu}\int \, d\tilde{\mu}^{(k)} \ k^{-3}
\exp\left(-kS_{0}+ik\sigma\right)
\label{totalm}
\end{eqnarray} 
where 
\begin{equation}
\int \, d\bar{\mu}=\frac{1}{2^{5}\pi M_{W}S_{0}^{2}}\int \,
d^{3}X \ 
\int \, \prod_{M=1}^{4} d^{2}\xi_{M}\,
\label{k=1}
\end{equation} 
and $d\tilde{\mu}^{(1)}=1$. For $k>1$,  
\begin{equation}
\int \, d\tilde{\mu}^{(k)} = \int \, \frac{ \prod_{q=1}^{4(k-1)}
dY^{q}\ d\alpha^{q}_{1} \ d\alpha^{q}_{2}}
{(2\pi)^{2(k-1)}\sqrt{{\rm det}\left(\tilde{g}\right)}} \ \exp \left(
-S^{(k)}_{\rm quad}\right)
\label{total2}
\end{equation}
\paragraph{}
As discussed above, in order to obtain a non-zero contribution, it is
necessary to saturate the Grassmann integrals in (\ref{k=1}) which
correspond to the eight unlifted zero modes protected by
supersymmetry. The simplest correlator which is non-zero in the
instanton background  must therefore have eight fermion
insertions. This part of the calculation is a straightforward
extension of the calculation given in \cite{dkmtv} of a four-fermion
correlator in the $N=4$ theory. Specifically we consider the instanton
contribution to a correlator with two insertions of each species of
adjoint fermion, 
\begin{equation}
G^{(8)}(x_{1},x_{2},\ldots,x_{8}) =  \langle \lambda_{\alpha}(x_{1})
\lambda_{\beta}(x_{2})\psi_{\gamma}(x_{3})\psi_{\delta}(x_{4})
f_{\epsilon}(x_{5})f_{\kappa}(x_{6})\tilde{f}_{\rho}(x_{7})
\tilde{f}_{\sigma}(x_{8}) \rangle 
\label{correlator}
\end{equation}
where, as in \cite{dkmtv}, we have defined low-energy massless fermion
fields by\footnote{Recall that the
$SU(4)_{\cal R}$ index $M$ 
labels the four species of Weyl fermions.} $\lambda^{M}_{\alpha}={\rm
Tr}(\ulambda^{M}_{\alpha}\tau^{3})$. At leading semiclassical order
each of these fields is replaced by its large-distance (LD) behaviour of
the corresponding zero modes in the monopole background:  
\begin{eqnarray}
(\lambda^{M}_{\alpha})^{\rm LD} & =  & 8\pi k 
\xi^{M}_{\beta} 
\S_{\rm F}(x-X)_{\alpha}^{\ \beta}
\label{ld}
\end{eqnarray} 
This asymptotic form is valid for 
$|x-X| \gg  M_{W}^{-1}$ where $X$ is the centre of mass of the $k$
instanton configuration and $\S_{\rm F}(x)=\gamma_\mu x_\mu/(4\pi
|x|^{2})$ is the three-dimensional Weyl fermion propagator. The factor
$k$ in (\ref{ld}) reflects the coefficient of the long-range magnetic Coulomb
fields of the charge-$k$ monopole.      
\paragraph{}
The final result can be expressed as a
contribution to an eight anti-fermion vertex which is a correction to
the classical low-energy effective action for the massless fermions. 
The latter is just a free massless action for the Weyl degrees of
freedom. 
\begin{equation}
S_{\rm F}=\frac{2\pi}{e^{2}}
\int\,d^{3}x \,
i\bar{\lambda}^{M}\bar{\sigma}_{\mu}\partial_{\mu}\lambda^{M}
\label{lclf}
\end{equation}     
where a sum over the $SU(4)_{\cal R}$ index is implied. The resulting
eight anti-fermion vertex can be written as a sum over contributions
from sectors of different (positive) topological charge $k$, 
\begin{equation}
S_{I}=\sum_{k=1}^{\infty} {\cal V}_{k}
\exp\left(-k S_{0} + ik\sigma\right) 
 \int \, d^{3}x \
\prod_{M=1}^{4} \bar{\lambda}^{2}_{M}
\label{vertex}
\end{equation}
Evaluating the integrals over $\xi^{M}_{\alpha}$ we obtain, 
\begin{eqnarray}
{\cal V}_{k} & = & \frac{{\cal V} k^{5}}{S_{0}^{3}} 
\int\, d\tilde{\mu}^{(k)}
\nonumber \\ 
{\cal V} & = & \frac{2^{14}\pi^{9}}{e^{2}} \,
\left(\frac{2\pi}
{e^{2}}\right)^{8}
\label{v1}
\end{eqnarray}
where the eight powers of $2\pi/e^{2}$ in (\ref{v1}) reflect  
our choice of normalization for the kinetic term in (\ref{lclf}).  
For $k=1$, after a comparison of normalizations and
definitions, we find agreement with the one-instanton result 
of Polchinski and
Pouliot \cite{polch} (their equation (3.23)). 
\paragraph{}
For $k>1$, the remaining problem is to evaluate the integral (\ref{total2}) 
over the parameters which correspond to the 
relative moduli of the $k$-monopole solution and their
superpartners. The fermionic integrals in this expression are
saturated by bringing down $2(k-1)$ powers of quadrilinear term
$S^{(k)}_{\rm quad}$. After performing the Grassmann integrations we are
left with a bosonic integral over the reduced charge-$k$
moduli space. Setting $d=4(k-1)$ we have,  
\begin{equation}
\int \, d\tilde{\mu}^{(k)} = \frac{1}{(8\pi)^{d/2}(d/2)!}
\int \, \frac{\prod_{q=1}^{d}
dY^{q}}
{\sqrt{{\rm det}\left(\tilde{g}\right)}}\ \varepsilon^{p_{1}p_{2}\ldots
p_{d}}\ \varepsilon^{q_{1}q_{2}\ldots
q_{d}}\ \tilde{R}_{p_{1}p_{2}q_{1}q_{2}}\ldots 
\tilde{R}_{p_{d-1}p_{d}q_{d-1}q_{d}}
\label{gbh}
\end{equation}
This integral has a familiar form: it is precisely the volume
contribution to the index of the Euler operator on the non-compact
manifold $\tilde{\cal M}_{k}$. If we were integrating over a compact
manifold it would simply be equal to the Euler characteristic of that 
manifold by the Gauss-Bonnet-Chern theorem. As usual, an extension to
the case of a non-compact manifold can be obtained from a limiting case
of the index theorem on a manifold with boundary\footnote{Index theory
on manifolds with boundary is reviewed in \cite{egh}.}. 
In the present case the
Euler character of the manifold can be writen as the sum of the volume term
(\ref{gbh}) and a surface term which involves the 
integral of a certain Chern-Simons form over the boundary. As we
discuss below, these surface terms are known to vanish for the case
$k=2$ and we will conjecture that they also vanish for all $k>2$.   
\paragraph{}
The centered moduli-space of two BPS monopoles has
real dimension four. In this case, the GBC integral has
been evaluated explicitly by Gauntlett and Harvey \cite{gh} 
using the known metric on 
$\tilde{\cal M}_{2}$. They obtained,  
\begin{equation}
\int \, d\tilde{\mu}^{(2)} = - 
\frac{1}{32\pi^{2}} \int_{\tilde{\cal M}_{2}} \ \varepsilon^{abcd} \,
\tilde{R}_{ab}\,\wedge\,{^*\tilde R_{cd}}  = 2
\label{ke2}
\end{equation}
where the integrand is written in a vierbein basis in terms of the 
curvature 2-forms $\tilde{R}_{ab}$. 
In fact, the Euler character $\chi(\tilde{\cal M}_{2})$ is also equal
to two, reflecting the fact that the manifold contracts onto a  
two-sphere which is the double-cover of the bolt. Hence, in this case,
the boundary terms in the index theorem must vanish in the infinite
volume limit where the boundary is removed to infinity. 
This is consistent with a previous analysis of Gibbons,
Pope and R\"{o}mer \cite{GPR}, 
who showed that the boundary contribution vanishes
for a class of metrics, known as Asymptotically Locally Flat (ALF)
 metrics, of which the
Atiyah-Hitchin metric on $\tilde{\cal M}_{2}$ is a particular example.  
\paragraph{}
Unfortunately, it is much harder to make a precise statement about the
boundary contributions to the index theorem for the case $k>2$. Here,
it is known that the metric approaches a multi-centre 
Taub-NUT metric in the limit in which the separation
between each pair of monopoles becomes large \cite{gm2}. This metric 
has the same 
asymptotic flatness properties as the two-monopole metric and it is
very plausible that the surface contributions from these parts of the
boundary vanish as they do for $k=2$. However, the `boundary at infinity'
of the $\tilde{\cal M}_{k}$ for $k>2$ also contain clustering regions
in which at least one pair of monopoles remain at finite  
separation. Heuristically one may argue that these regions represent a
fraction of the boundary which tends to zero in the infinite volume
limit. As the metric and therefore the surface-integrand 
is known to be non-singular, 
it should be possible show that the contribution of
these regions to
the surface integral is bounded above by some inverse power of the
volume. In the absence of a more precise analysis of this issue we will
simply assume that 
the surface contribution does vanish in this limit and therefore that the
GBC integral (\ref{gbh}) is equal to the Euler
character $\chi(\tilde{\cal M}_{k})$ for all $k$.               
\paragraph{}
The cohomology of the centered multi-monopole moduli spaces, 
$\tilde{\cal M}_{k}$, 
has recently been determined from description of these spaces as
spaces of rational maps \cite{segal}. 
The cohomology with complex coefficients, 
$H^{*}(\tilde{\cal M}_{k})$, is divided into different sectors
according to the action of 
the discrete symmetry group, ${\bigZ}_{k}$. Let $H^{*}(\tilde{\cal
M}_{k})_{p}$ denote the part of the cohomology where $\xi \in 
{\bigZ}_{k}$ acts as
$\xi \rightarrow \xi^{p}$. Then the result of Ref.~\cite{segal} 
is that $H^{i}(\tilde{\cal M}_{k})_{p}$ 
has complex dimension one whenever 
$i=2k-2(p,k)$ where $(p,k)$ is the greatest common divisor of $p$ and
$k$, and dimension zero otherwise. As the non-vanishing cohomology is even,
the Euler character is obtained by counting the total number of
solutions of the condition $i=2k-2(p,k)$; 
\begin{equation}
\chi(\tilde{\cal M}_{k})=\sum_{i=0}^{4(k-1)} \ \sum_{p=0}^{k-1} 
\delta_{i,2k-(p,q)} = k       
\label{result}
\end{equation}
Hence our final result for the multi-instanton contribution to the
vertex (\ref{vertex}) is ${\cal V}_{k}=k^{6}{\cal V}/S_{0}^{3}$. Now the
instanton series may be summed straightforwardly to give; 
\begin{equation} 
S_{I}= {\cal V}
\left[\frac{1}{2S_{0}^{3}}\left(\frac{\partial}{\partial S_{0}}\right)^{6} \, 
\coth\left(\frac{S_{0}}{2}+ \frac{i\sigma}{2}\right)\right] 
 \int \, d^{3}x \
\prod_{M=1}^{4} \bar{\lambda}^{2}_{M} 
\label{vertex2}
\end{equation}
Including the effects of multi-instantons of negative topological
charge yields a vertex for the left-handed Weyl fermions. The result
is obtained by making the replacements, 
$\bar{\lambda}_{M}^{2}\rightarrow {\lambda}_{M}^{2}$ and 
$\sigma \rightarrow -\sigma$ in (\ref{vertex2}). 
\paragraph{}
We finish by comparing the resummed multi-instanton vertex obtained
above with the amplitude for membrane scattering in eleven-dimensional
supergravity calculated in \cite{polch}. The amplitude given in 
equation (3.39) of this reference 
yields a prediction for a term, $S_{4\hbox{-}\partial}$, in the low-energy
effective action of the 3D SUSY gauge theory which has four time
derivatives. In our notation this term involves four powers of 
$\partial_{0}v_{3}$, where $v_{3}={\rm Tr}(\uv_{3}\tau^{3})$ is the
only low-energy scalar field (other than $\sigma$) with a non-zero
VEV, 
\begin{equation}
S_{4\hbox{-}\partial}= K \int \, d^{3}x \left(\partial_{0}v_{3}\right)^{4} \ 
\sum_{n=-\infty}^{+\infty} 
\frac{1}{\left(S_{0}^{2}+(2\pi n-\sigma)^{2}\right)^{3}}
\label{sugra}
\end{equation}
Here $K$ is an overall constant which is independent of the VEV. 
We will evaluate the sum on the RHS
of (\ref{sugra}) using the identity, 
\begin{equation}
\sum_{n=-\infty}^{+\infty} 
\frac{1}{S_{0}^{2}+(2\pi n-\sigma)^{2}} = 
\frac{1}{4S_{0}}\left[ \coth\left(\frac{S_{0}}{2}+
\frac{i\sigma}{2}\right) + 
\coth\left(\frac{S_{0}}{2}- \frac{i\sigma}{2}\right)\right]
\label{identity}
\end{equation}
The required sum is obtained by differentiating the above relation
twice with respect to $S_{0}^{2}$. The presence of 
four-derivative term (\ref{sugra}) can be checked directly in the instanton
approach by saturating the eight Grassmann integrals in (\ref{k=1})
with four insertions of the fermion bilinear terms in the scalar field
$v_{3}$. Equivalently, we may compare the instanton-induced 
vertex (\ref{vertex2}) with the eight-fermion vertex which 
arises as the supersymmetric completion of (\ref{sugra}) in the
low-energy effective action. Up to a constant, 
the coefficient of the eight-fermion vertex is obtained from that of the 
four-derivative term by differentiating the latter 
with respect to the VEV four times\footnote{This is analogous to the
relation between the instanton
contributions to the kinetic terms and to the four-fermion term in the
four-dimensional theories studied in \cite{dkm1}.}. 
The vertex for the right-handed fermions obtained in this way can be
written as,     
\begin{equation}
\tilde{S}_{I}= 2{\cal V}
\left[\left(\frac{\partial}{\partial S_{0}}\right)^{4} 
\left(\frac{1}{2S_{0}} \frac{\partial}{\partial S_{0}}\right)^{2}  
\frac{1}{S_{0}}
\coth\left(\frac{S_{0}}{2}+ \frac{i\sigma}{2}\right)\right] 
 \int \, d^{3}x \
\prod_{M=1}^{4} \bar{\lambda}^{2}_{M} 
\label{vertex3}
\end{equation}
The predicted vertex has a double expansion in powers of $1/S_{0} \sim
e^{2}/{\rm v}$ and
in powers of the instanton action $\exp(-S_{0}+i\sigma)$. The latter
expansion is an expansion in instanton number and, to compare with 
multi-instanton result (\ref{vertex2}) we will retain
terms of all orders.  The former expansion corresponds to 
a series of perturbative corrections and, as
we wish to compare with a leading order semiclassical calculation, we
must retain only the leading order term in each instanton sector. 
To this order the partial derivatives in (\ref{vertex3}) can all be moved
to the right to act on the hyperbolic cotangent and we find agreement
with (\ref{vertex2}). 
\paragraph{}
\centerline{******************}

The authors would like to thank C. Fraser, J. Gauntlett, G. Gibbons, 
N. Manton, G. Segal and D. Tong for useful discussions. 
ND and VVK would like to thank the Isaac 
Newton Institute for hospitality while part of this work was completed.  
ND is supported by a PPARC Advanced Research Fellowship. 
MPM is supported by the Department of Energy.

\end{document}